\documentclass[prl,aps,twocolumn,showpacs,superscriptaddress,dvips,linenumbers]{revtex4}
\usepackage{amsmath}
\usepackage{amssymb}
\usepackage{graphicx}
\usepackage{graphics}
\usepackage{dcolumn}
\usepackage{bm}

\begin{document}

\title{Spin-thermodynamics of cold spin-$1$ atoms decoupled from spatial modes}
\author{Z. B. Li}
\author{D. X. Yao}
\author{C. G. Bao}
\email{stsbcg@mail.sysu.edu.cn}

\begin{abstract}
We study the thermodynamic properties of cold spin-1 atoms with a
fixed magnetization $M$ and decoupled from spatial modes. Three
temperature domains are found: ($0$, $T_{1}$) is a domain of
second condensation, namely, both the spatial and spin degrees of
freedom are frozen; ($T_{1}$, $T_{2}$) is a $T$-sensitive domain,
where the internal energy $U \propto T$, entropy $S_{E} \propto \log T$, and $U$/$k_{B}T$ is always less than $3$/$2$; % a upper limit $U \le \frac{3}{2}k_{B}T$ is found;
 ($T_{2}$, $T_{3}$) is the third domain with a maximum entropy.
When $T$ is higher than $T_{3}$, the spatial modes can not be
neglected. The appearance of these domains originates from the two
gaps: (i) The gap between the ground state and the first excited
state, and (ii) the gap between the highest spin-state without
spatial excitation and the lowest state with a spatial mode excited.
These two gaps are crucial to the low temperature physics and they
can be tuned.

\end{abstract}

\maketitle

\address{The State Key Laboratory of Optoelectronic Materials and
Technologies \\ School of Physics and Engineering, Sun Yat-Sen
University, Guangzhou, P. R. China}

\bigskip Since the pioneering experiment on spinor condensation \cite{STE98}, the
study of these artificial and tunable systems becomes a hot topic
\cite{STE98,ho98,ohmi98,law98,pu99,zhan09,chan05,kawa12}. Due to the
rapid development of low-temperature techniques
\cite{jing11,pasq11,paz12,pasq12}, a magnificent goal is to explore
the domain below the spatial condensation temperature $T_{c}$. In
this domain, the condensation can be regarded as a pure spin-system
with the frozen spatial modes. Furthermore, when the particle number
$N$ is fixed and the magnetization $M$ is conserved, the dimension
of the spin-space is given by $(N-M+2)/2$. Thus the dimension is
tunable and could be very low. In addition to the theoretical
interest, the low dimensional spin-systems are very promising for
important applications. For example, a two-dimensional system could
be used as a quantum bit \cite{conn10,walt13}. Furthermore, a recent
paper shows that the spin oscillation of a highly multi-spatial-mode
thermodynamical gas can be described by a theory without considering
the spatial degrees of freedom \cite{Pech}. It implies that, the
spin-mode can be decoupled from the spatial mode under certain
conditions. Therefore, the study of a pure spin-system is
meaningful. The purpose of this paper is to (i) explore the
thermodynamic properties \cite{yama82} of a pure spin-system and the
way to measure them, and (ii) find out the conditions that a
condensate can be considered as a pure spin-system.

Consider $N$ spin-$1$ bosons trapped by an isotropic harmonic
potential (HP) $\frac{1}{2}m\omega^{2}r^{2}$. Firstly, the magnetic field
$B$ is assumed to be zero. Then, the effect of a residual $B$ is
studied. When the dipole-dipole interaction is negligible and $B=0$,
both the total spin $S$ and $M$ are conserved. When the spatial
excitation is not involved, the eigenstates can be written as $\Psi_{SM}=\Pi_{i}\phi_{S}(\mathbf{r}%
_{i})\vartheta_{SM}^{N}$, where $\phi_{S}(\mathbf{r}_{i})$\ can be
obtained by solving the generalized Gross-Pitaevskii equation
conserving both $S$ and $M$ \cite{bao04}, $\vartheta_{SM}^{N}$\ is
the normalized all-symmetric spin-state with good
quantum numbers $S$ and $M$. This spin-state is unique for a pair of ($S$%
,$M$), and $N-S$ must be even due to the boson statistics \cite{katr01,bao04}.
The eigenenergy relative to the ground state (g.s.) is
\begin{equation}
E_{S}=\frac{c_{S}}{2}[S(S+1)-S_{g}(S_{g}+1)], \label{eq1}%
\end{equation}
where $c_{S}=c_{2}\int|\phi_{S}(\mathbf{r})|^{4}d\mathbf{r}$, $c_{2}$ is the
strength of the spin-dependent interaction. It was found that $|(c_{S}%
-c_{S^{\prime}})/c_{S}|<0.02$\ in relevant cases. Therefore the
subscripts in $c_{S}$ and $\phi_{S}$ can be dropped. $S_{g}$
is the total spin of the g.s., we have $S_{g}=N$ (if $c_{2}<0$, say,
$^{87}$Rb) or $=M^{\prime}$ (if $c_{2}>0$, say, $^{23}$Na), where
$M^{\prime}=M ($or $M+1)$ if $N-M$ is even (or odd), and $M\geq0$
is assumed.

When the system approaches the thermodynamic equilibrium, it can be treated as a
canonical ensemble. The partition function is $Z(c,T)\equiv\Sigma_{S}^{\prime
}\ e^{-\beta E_{S}}$, where $S$ is from $M^{\prime}$ to $N$ with even $N-S$,
 and $\beta=1/k_{B}T$. The internal energy $U(c,T)=-\frac
{\partial}{\partial\beta}\ln Z$, the specific heat $C%
(c,T)=\frac{\partial U}{\partial T}$, and the entropy $S_E%
(c,T)=k_{B}\ln Z+U/T$. Since $c$ is adjustable, the specific
strength $Y(c,T)\equiv\frac{\partial U}{\partial c}$ is further
defined to connect other state functions.

One can prove that $U$\ satisfies the partial differential equation as
\begin{equation}
(T\frac{\partial}{\partial T}+c\frac{\partial}{\partial c}-1)U(c,T)=0.
\label{eq2}%
\end{equation}
The other three functions satisfy
\begin{equation}
T\frac{\partial F}{\partial T}+c\frac{\partial F}{\partial c}=0, \label{eq3}%
\end{equation}
where $F$ represents $C$, $Y$, or $S_E$. They are subjected to the
boundary conditions listed in Table I. When $T\rightarrow\infty$ or
$c\rightarrow0$, all the weights $e^{-\beta E_{S}}\rightarrow1$.
Thus, $Z\rightarrow N_{state}$, where $N_{state}=(N-M^{\prime}+2)/2$
is the dimension of the spin-space. In the same limit,
$U\rightarrow|cN_{U}|$, $N_{U}=-(N-M^{\prime})(4N+2M^{\prime}+1)/12$
(if $c<0$) or $(N-M^{\prime })(2N+4M^{\prime}+5)/12$ (if $c>0$),
which is proportional to the direct sum of the eigenenergies, and
$U/T$\ will tend to zero. Therefore, the entropy $S_E /
k_{B}\rightarrow\ln Z=\ln N_{state}$ as shown in Table I. Whereas
when $T\rightarrow0$ or $|c|\rightarrow\infty$, the weight of the
g.s. is one while all the other weights become zero. Since the g.s.
is not degenerate, $S_E$ vanishes in this limit.

\begin{table}[pb]
\caption{The boundary conditions for thermodynamic functions.}%
\label{table.1}
\begin{center}%
\begin{tabular}
[c]{ccccccccc}%
~ &  & $U$ &  & $C$ &  & $Y$ &  & $S_E$/$k_{B}$\\
$T\rightarrow0$ &  & $0$ &  & $0$ &  & $0$ &  & $0$\\
$T\rightarrow\infty$ &  & $|cN_{U}|$ &  & $0$ &  & $N_{U}$ &  & $\ln
N_{state}$\\
$c\rightarrow0$ &  & $0$ &  & $0$ &  & $N_{U}$ &  & $\ln N_{state}$\\
$c\rightarrow\infty$ &  & $0$ &  & $0$ &  & $0$ &  & $0$%
\end{tabular}
\end{center}
\end{table}

Additionally, $U$ satisfies $T^{2}\frac{\partial^{2}U}{\partial T^{2}}-c^{2}\frac{\partial^{2}U}{\partial c^{2}}=0$. From this equation, we obtain
a relation between $C$ and $Y$ as $T\frac{\partial C
}{\partial T}=-c\frac{\partial Y}{\partial T}$ and $T\frac{\partial C}{\partial c}=-c\frac{\partial Y}{\partial c}$. They imply that
$C$ and $Y$ can vary simultaneously with $T$ and/or $c$ if $c<0$,
or they vary oppositely if $c>0$. Moreover, we have $\frac
{\partial S_E}{\partial T}=C/T$ and $\frac{\partial S_E}{\partial c}=-C/c$. Thus the increasing of $T$\ always
leads to an increase of $S_E$, while the increasing of $|c|$\ always
leads to a decrease of $S_E$. In particular, Eq. (2) relates the three
thermodynamic functions to each other, $U=TC+cY$.

Numerical examples are shown in Fig. 1, where the realistic
parameters of interaction are used. The variation of thermodynamic
functions versus $T$ shows three steps: insensitive, sensitive, and
insensitive again. For example, the $U$ of Rb and Na remains to be
zero when $T$ is very low. Once $T$\ is higher than a turning point
$T_{1}$, a sudden increasing of $U$\ happens (as shown in Fig.1a).
When $T$\ is higher than the second turning point $T_{2}$, $U$ does
not increase any more and remains $\approx|cN_{U}|$. Note that the
energy gap between the g.s. and the first excited state is
$E_{gap,1}=|c|(2N-1)$ (if $c<0$) or $=|c|(2M^{\prime}+3)$ (if
$c>0$). Thus, when $M^{\prime}$ is small, the gap for Rb is much
larger than the gap for Na. On the other hand, we find $T_{1}$ for
Rb is much larger than the one for Na from Figs. 1a and 1e.
Therefore, a larger gap might lead to a higher $T_{1}$. Furthermore,
for Na, among the three curves in Fig. 1e with different $M$, the
dotted curve with $M=1500$ has the largest $T_{1}$. It suggests also
that a larger gap might give a higher $T_{1}$.

The domain ($T_{1},T_{2}$) is a $T-$sensitive domain, where $U$\
increases almost linearly with $T$ (accordingly $C$ is
close to a constant), and also $S_E$\ increases nearly
linearly with $\ln T$ disregarding $c<0$ or $c>0$. When $c<0$, we
know from Fig. 1b and 1c that the variations of $C$ and $Y$
are synchronous as predicted. Whereas they behave Oppositely when
$c>0$ as shown in Figs. 1f and 1g. It is worth to point out that the lower
part of the spectrum depends (does not depend) on $M$ if $c>0$ ($<$ 0).
Therefore, all the curves in Fig. 1 for Na depend on $M$\ strongly,
while those for Rb do not.

In the g.s., both the spatial and spin degrees of freedom are
frozen. This can be called the second condensation \cite{pasq12}.
The probabilities at the g.s. is $P_{g}=1/Z$. When $T=0$, $P_{g}=1$.
When $P_{g}$\ decreases from 1, the spin-fluctuation begins. The
probability at the highest spin-state is $P_{top}=e^{-\beta
E_{top}}\ /Z$, where
$E_{top}=\frac{|c|}{2}[N(N+1)-M^{\prime}(M^{\prime}+1)]$ is the
energy difference between the highest and the lowest spin-states.
Note that, in the pure spin-space,
$\underset{T\rightarrow\infty}{\lim}P_{top}\rightarrow 1/N_{state}$.
Thus the deviation of $P_{top}N_{state}$ from 1 measures how far the
spin-fluctuation is away from the maximum value. Examples of $P_{g}$
and $P_{top}N_{state}$ are shown in Fig. 2. To show the
$T-$sensitivity, $U/k_{B}T$ and $\mathfrak{C/}k_{B}$ are also
plotted. It is found that $T_{1}$\ is the position that $P_{g}$
starts to decline from 1. Thus $T_{1}$ marks the temperature of the
second condensation. $T_{2}$\ is the place that $P_{top}N_{state}$
starts to decrease from 1. Thus $T_{2}$ marks the maxima of
spin-fluctuation and entropy. To describe it quantitatively,
$T_{1}$\ and $T_{2}$ are defined at which $P_{g}=0.95$ and\
$P_{top}N_{state}=0.95$, respectively. When $T$\ is very low,
$P_{g}$ can be approximated by $P_{g,appr}=1/(1+e^{-\beta
E_{gap,1}})$, and accordingly $T_{1}=0.34\ E_{gap,1}/k_{B}$.

Furthermore, our calculation demonstrates that $U/k_{B}T$ is always less than $3/2$ as shown in Fig. 2. It implies that, disregarding $N$\ and
interactions, the internal energy contributed from all the spin degrees of
freedom is even smaller than the energy assigned to the spatial motion for a single particle.
 This fact manifests how small the energy is involved in the spin-space.

Let the energy of the lowest state with a spatial mode excited be
$E^{ex}_{1}$, which can be obtained by solving the equation given in
the ref.\cite{bao05,pang} How weak the interference from the spatial
mode would be depends on the gap $E_{gap2}=E^{ex}_{1}-E_{top}$. When
the gap is sufficiently large, the probability of staying in the
spatially excited levels is negligible. The temperature at which
$e^{-\beta E_{gap,2}}=0.05$ is defined as $T_{3}$ (i.e.,
$T_{3}=\frac{E_{gap,2}}{3k_{B}}$). When $T<T_{3}$, the spin-space is
decoupled from the spatial modes.

When both $N$ and $M$ are fixed, the dependence of $T_{1,}$,
$T_{2}$, and $T_{3}$ on $\omega$ is shown in Fig. 3. Note that a
larger $\omega$\ will lead to a more compact $\phi(\mathbf{r)}$
(smaller in size) and thereby a stronger $|c|$. Hence, all the $T_1,
T_2$ and $T_3$ will become higher when $\omega$ increases. For Rb,
the second condensation can be realized at $T_{1}=10^{-9}K$ when
$\omega =10^{4.5}s^{-1}$ and $N=1000$(Fig.3a). If $\omega $ and/or
$N$ are larger, $T_1$ is even higher. For Na, $E_{gap1} \propto
M^{\prime}$, therefore $T_1$ depends on $M$ seriously. When $M=N-2$
and $N=1000$, $T_{1}$ would be higher than $10^{-9}K$ if $\omega >
10^{3.7}s^{-1}$ (3c).

When $N$ is small $T_3$ can be quite high (say, $T_3
>10^{-8}K$ if $\omega > 10^4 s^{-1}$ (3b,3d). It implies that a nearly pure spin-system can be
created experimentally. When $N$ is larger, due to the cross over of
the highest member of the ground band and the lowest member of the
excited band, $T_3 $ does not exist unless $M$ is close to $N$ (3a
and 3c). In 3a (3c) and when $\omega =10^{3.1} s^{-1} (10^{2.28}
s^{-1})$, a cross over of $T_3$ and $T_2$ occurs. When $\omega $ is
smaller than the value, the domain ($T_1,T_2$) is under $T_3$ so
that the whole process of the increase of entropy from zero to being
maximized is free from the interference of the spatial modes.

 Since the thermodynamic functions change greatly in the $T-$sensitive
domain, the spin-texture and spin-component $\mu$ should be
modulated accordingly. For $\vartheta_{SM}^{N}$, the probability of
a particle in $\mu$ can be calculated from Eq. (10) in Ref.
\cite{bao06}. In particular, when $\mu=0$ the probability is
\begin{equation}
P_{0}^{SM}=\frac{S(S+1)(2N+1)-M^{2}(2N+3)-N}{N(2S+3)(2S-1)}
\end{equation}
When the thermo-fluctuation is taken into account, we define
\begin{equation}
\overline{P}_{0}^{M}(T)=\frac{1}{Z}\Sigma _{S}^{\prime }\
P_{0}^{SM}e^{-\beta E_{S}}
\end{equation}
and the population of $\mu=0$ component is
$N\overline{P}_{0}^{M}(T)$. Examples of $\overline{P}_{0}^{M}(T)$\
are given in Fig. 4. When $T\rightarrow0$ and for Rb, $\overline{P}_{0}^{M}(0)$=$P_{0}^{NM}$%
=$\frac{N^{2}-M^{2}}{N(2N-1)}$, while for Na, $\overline{P}_{0}^{M}%
(0)$=$P_{0}^{MM}$=$\frac{N-M}{N(2M+3)}$. In both cases,
$\overline{P}_{0}^{M}(0) $ decreases with $M$. As $T$ increases,
$\overline{P}_{0}^{M}(T)$\ remains unchanged until $T$ enters into
the $T-$sensitive domain. Thus the borders of the $T-$sensitive
domain can be evaluated by measuring $\overline{P}_{0}^{M}(T)$.

When a magnetic field $B$ is applied, the linear Zeeman term can be
dropped for the $M-$conserved systems, the quadratic Zeeman term
corresponds to $H_{B}=q\sum _{i}\mathbf{f}_{iz}^{2}$, where
$\mathbf{f}_{iz}$ is the z-component of the spin-operator for the
$i$-th particle. Under $H_{B}$\, the states with different $S$ will
mix up, and the $i$-th eigenstates will appear as
$\Psi_{i,M}^{B}=\sum _{S}^{^{\prime}}C_{S}^{B,i}\Psi_{SM}$. The
matrix elements $\langle\vartheta_{S^{\prime}M}^{N}|H_{B}|\vartheta_{SM}%
^{N}\rangle$\ are shown in Eqs. (3,4,6,7) of Ref. \cite{bao06}. By
diagonalizing the matrix of $H+H_{B}$, the eigenenergies
$E_{i}^{B}$\ and $\Psi_{i,M}^{B}$ can be obtained. It is found that
there is always a turning point $B_{0}$ so that the spectra will
remain nearly unchanged when $B\leq B_{0}$ as shown in Fig. 5, where
$B_{0}=0.5mG$. Based on the perturbation theory
$\Psi_{i,M}^{B+\varepsilon}$ would be closer to $\Psi_{i,M}^{B}$
(where $\varepsilon$\ is a small quantity) if the levels in the
neighborhood of $E_{i}^{B}$ are wider separated. Accordingly, the
$B_{0}$ for Rb is much larger than that for Na because the low-lying
levels of the former are much splitting Furthermore, a smaller $N$\
can lead to a larger $\int|\phi(\mathbf{r})|^{4}d\mathbf{r}$,
therefore a larger $|c|$ and accordingly a larger level-separation
resulting also in a larger $B_{0}$. For Na, the level-separation
depends on $\mathbf{M}$, hence a larger $M$ will also lead to a
larger $B_{0}$. Since all the thermodynamic properties depend solely
on the spectra, the invariance of the spectra implies that all the
features at $B=0$ will remain unchanged when $B\leq B_{0}$.

In summary, the $M-$conserved pure spin-systems of cold Rb and Na atoms at
$B=0$ are studied, three domains of $T$ are found. The effect of a residual
$B$ has been evaluated. The main results are:

1) ($0$, $T_{1}$) is a $T-insensitive$ domain originated from the
gap $E_{gap,1}$, and $T_{1}=0.34\ E_{gap,1}/k_{B}$ marks the
\textit{second condensation } \cite{pasq12}. It is reported (Fig. 2
of Ref. \cite{pasq12}) that, for a spin-$3$ condensation, $M$
depends on $T$ in general, but becomes insensitive to $T$ when
$T\rightarrow0$ and $B$ is sufficiently weak. Thus, the appearance
of the $T-insensitive$ domain during $T\rightarrow 0$\ might be a
common phenomenon.

2) $(T_{1}, T_{2})$ is a $T-$sensitive domain, where the
thermo-fluctuation keeps strengthening. $U \propto T$ and $S_E
\propto\ln T$ roughly hold, and $U/k_{B}T<\frac{3}{2}$ holds. The
location of this domain can be known by measuring
$\overline{P}_{0}^{M}(T)$.

3) $(T_{2}, T_{3})$ is again a $T-$insensitive domain with
$T_{3}=\frac{E_{gap,2}}{3k_{B}}$. Due to the gap $E_{gap,2}$, the
excitation of spatial modes is suppressed.

4) Both $E_{gap,1}$ and $E_{gap,2}$ can be tuned by changing
$\omega$, $N$, and $M$. Correspondingly, $T_{1}$, $T_{2}$, and
$T_{3}$ also change. A larger $N$ (for Rb) or a larger $|M|$\ (for
Na) will lead to a larger $E_{gap,1}$ and the second condensation
can be realized at higher temperature.  The domain $(T_{1}, T_{2})$
can be compressed by reducing
 $E_{top}$ (i.e., by increasing $M$).In particular, the low-dimensional systems are notable
for application. For example, a 2-level system can be formed by
fixing $M=N-2$.

5) The present techniques can provide a shield so that the effect of a
residual $B$ is negligible. On the other hand, when $B$ is stronger, its effect should be studied further.

6) It is also possible to consider an optical-lattice (OL) potential, or maybe an HP-OL combination.
 In that case, a deep OL may additionally freeze the spatial mode to realize the pure
spin-system.

As a final remarks, the two gaps $E_{gap,1}$ and $E_{gap,2}$ are in
general crucial to the physics at $T<T_{c}$.

Acknowledgment: The project is supported by the National Basic Research
Program of China (2008AA03A314, 2012CB821400), NSFC projects (11274393,
11074310, 11275279), RFDPHE of China (20110171110026) and NCET-11-0547.

\newpage

\begin{figure}[tbp]
\includegraphics{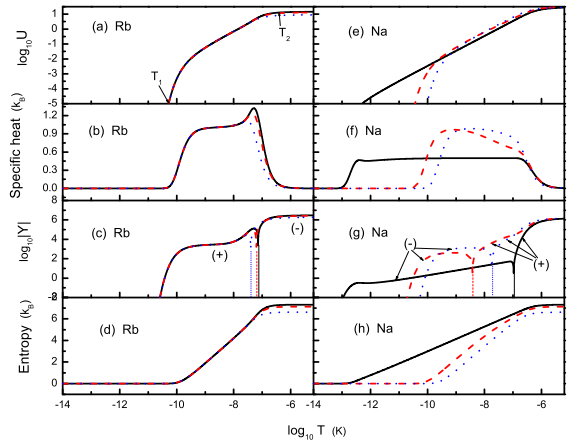}% Here is how to import EPS art
\caption{(Color online). $\log U$, $C$, $\log |Y|$, and $S_E$ of Rb
(left column) and Na (right column) condensates versus $\log_{10}
T$. The unit of $U$\ is $\hbar\omega$.  $N=3000$ and
$\omega=300\times 2\pi$ are assumed. $M$ is given at $0$ (solid
line), $500$ (dash), and $1500$ (dot). For (c) and (e), the sign of
$Y$ is marked by the curves.} \label{fig:1}
\end{figure}

\begin{figure}[tbp]
\includegraphics{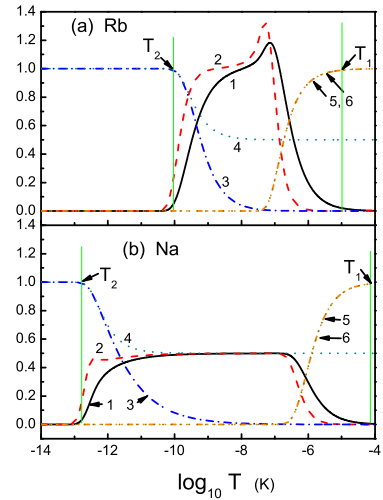}% Here is how to import EPS art
\caption{(Color online). Six functions of Rb (a) and Na (b) versus
$\log _{10}T$. Curve ``1" is for $U/k_{B}T$, ``2" is for $C/k_{B}$,
``3" is for $P_{g}$, ``4" is for $P_{g,appr}$, ``5" is for
$P_{top}N_{state}$, and ``6" is for $P_{top,ap}N_{state}$.
($P_{top,ap}$ is obtained by transforming the summation involved
into
an integration). $N=3000$, $M=0$, and $\omega=300\times2\pi$ are used.}%
\label{fig:2}
\end{figure}

\begin{figure}[tbp]
\includegraphics{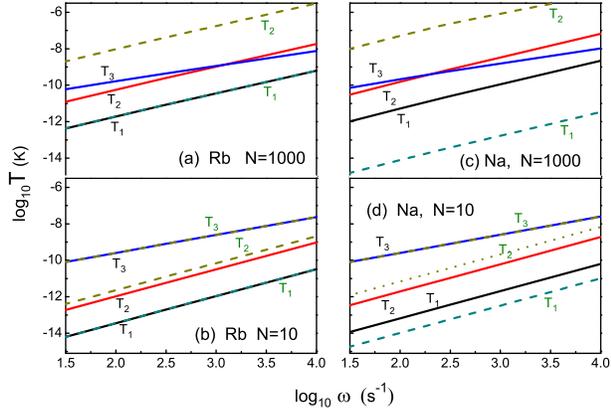}% Here is how to import EPS art
\caption{(Color online). $T_1, T_2$ and $T_3$ versus $\omega $.
Solid lines are for $M=N-2$, dash for $M=0$. When $M=0$ and
$N=1000$, $T_3$ does not exist.}%
\label{fig:3}
\end{figure}

\begin{figure}[tbp]
\includegraphics{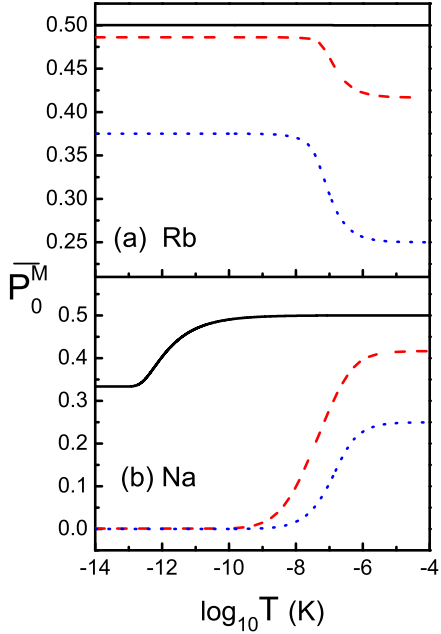}% Here is how to import EPS art
\caption{(Color online). $\overline{P}_{0}^{M}(T)$ versus $T$ for Rb
(a) and Na (b). The parameters and the three choices of $M$ are the
same as in Fig. 1. } \label{fig:4}
\end{figure}

\begin{figure}[tbp]
\includegraphics{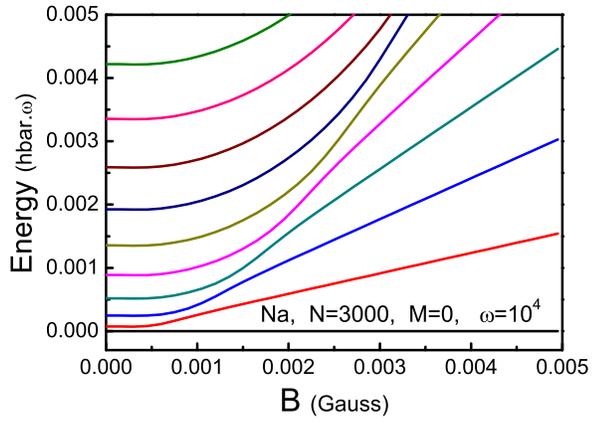}% Here is how to import EPS art
\caption{(Color online). The spectrum of a Na condensate versus $B$.
The lowest ten curves are plotted.} \label{fig:5}
\end{figure}

\end{document}